\documentclass{article}
\usepackage{spconf,amsmath,graphicx,amsfonts}
\usepackage{amsmath}
\usepackage{amssymb}

\usepackage{footnote}
\makesavenoteenv{tabular}
\usepackage{algorithm}
\usepackage{algorithmic}

\usepackage{mleftright}
\usepackage{epsfig,epstopdf}
\usepackage{adjustbox}
\usepackage{caption}
\ninept
\begin{document}

\title{Study of Compressed Randomized UTV Decompositions for Low-Rank Matrix Approximations in Data Science}
\name{\large{Maboud F. Kaloorazi  ~and Rodrigo C. de Lamare
\vspace{-0.5em}} }

%\address{{Centre for Telecommunications Studies (CETUC)}\\
%Pontifical Catholic University of Rio de Janeiro,Brazil\\
%Department of Electronics, University of York, United Kingdom\\
%Emails: {\kaloorazi,~delamare\}@cetuc.puc-rio.br}}

\address{\large {Centre for Telecommunications Studies (CETUC)}\\{\large Pontifical Catholic University of Rio de Janeiro, Brazil}\\
{\large Department of Electronics, University of York, United Kingdom}\\
{\large Emails: {\{kaloorazi,~delamare\}}@cetuc.puc-rio.br}
\vspace{-0.5em}}

\maketitle

\begin{abstract}
In this work, a novel rank-revealing matrix decomposition algorithm
termed Compressed Randomized UTV (CoR-UTV) decomposition along with
a CoR-UTV variant aided by the power method technique is proposed.
CoR-UTV computes an approximation to a low-rank input matrix by
making use of random sampling schemes. Given a large and dense
matrix of size $m\times n$ with numerical rank $k$, where $k \ll
\text{min} \{m,n\}$, CoR-UTV requires a few passes over the data,
and runs in $O(mnk)$ floating-point operations. Furthermore, CoR-UTV
can exploit modern computational platforms and can be optimized for
maximum efficiency. CoR-UTV is also applied for solving robust
principal component analysis problems. Simulations show that CoR-UTV
outperform existing approaches.
\end{abstract}

\maketitle
%\IEEEdisplaynontitleabstractindextext
% \fi
%\IEEEpeerreviewmaketitle

\section{Introduction}
\label{sec:intr}

Low-rank matrix approximations play an increasingly important role
in signal processing and its applications. Such compact
representations which retain the key features of a high-dimensional
matrix provide a significant reduction in memory requirements, and
more importantly, computational costs when the latter scales, e.g.,
according to a high-degree polynomial, with the dimensionality.
Matrices with low-rank structures have found many applications in
background subtraction \cite{8425659, KaDeDSP17}, system
identification \cite{FazelPST13}, IP network anomaly detection
\cite{KaDeICASSP17,NHussainP18}, latent variable graphical modeling
\cite{Chandrasekaran12}, subspace clustering \cite{RahmaniAtiaCoP17,
Oh2017} and sensor and multichannel signal processing
\cite{ifir2005,intadap2005,jio2007,jiomvdr2008,mwfccm2008,DeSa2009,jiols2010},
\cite{ccmjio2010,sjidf2010,jiomimo2011,jiostap2011,barc2011,uwbccm2011,wlmwf2012,wljio2014,dfjio2014,rdrb2015,mserjidf2015,dfalrd2016}.
.

Singular value decomposition (SVD) \cite{GolubVanLoan96} and the
rank-revealing QR (RRQR) decomposition \cite{Chan87, GuEisenstat96}
are among the most commonly used algorithms for computing a low-rank
approximation of a matrix. On the other hand, a UTV decomposition
\cite{Stewart98} is a compromise between the SVD and the RRQR
decomposition with the virtues of both: UTV (i) is more efficient
than the SVD, and (ii) provides information on the numerical null
space of the matrix \cite{Stewart98}. Given a matrix $\bf A$, the
UTV algorithm computes a decomposition ${\bf A =UTV}^T$, where ${\bf
U}$ and ${\bf V}$ have orthonormal columns, and ${\bf T}$ is
triangular (either lower or upper triangular). These deterministic
algorithms, however, are computationally expensive for large data
sets. Furthermore, standard techniques for their computation are
challenging to parallelize in order to utilize advanced computer
architectures \cite{HMT2009, Gu2015}. Recently developed algorithms
for low-rank approximations based on random sampling schemes,
however, have been shown to be remarkably efficient, highly accurate
and robust, and are known to outperform existing algorithms in many
practical situations \cite{FriezeKVS04, Rokhlin09, HMT2009,Gu2015}.
The power of randomized algorithms lies in that (i) they are
computationally efficient, and (ii) their main operations can be
optimized for maximum efficiency on modern architectures.

%\subsection{Contributions}

This work presents a novel randomized rank-revealing method termed
compressed randomized UTV (CoR-UTV) decomposition. Given a large and
dense rank-$k$ matrix ${\bf A} \in \mathbb R^{m \times n}$, CoR-UTV
computes a low-rank approximation $\hat{\bf A}_\text{CoR}$ of $\bf
A$ such that
\begin{equation}
\hat{\bf A}_\text{CoR}={\bf UTV}^T,
\label{eq_contri}
\end{equation}
where ${\bf U}$ and ${\bf V}$ have orthonormal columns, and ${\bf
T}$ is triangular (either lower or upper, whichever is preferred).
CoR-UTV only requires a few passes through data, for a matrix stored
externally, and runs in $O(mnk)$ floating-point operations (flops).
The operations of CoR-UTV involve matrix-matrix multiplication, the
QR and RRQR decompositions. Due to recently developed
Communication-Avoiding QR algorithms \cite{DemmGHL12,DemGGX15,
DuerschGu2017}, which can perform the computations with
optimal/minimum communication costs, CoR-UTV can be optimized for
peak machine performance on modern architectures. We illustrate,
through numerical examples, that CoR-UTV is rank-revealer and
provides a highly accurate low-rank approximation to a given matrix.
Furthermore, we apply CoR-UTV to solve the robust principal
component analysis (robust PCA) problem \cite{CSPW2009, CLMW2009},
i.e., to decompose a given matrix with grossly corrupted entries
into a low-rank matrix plus a sparse matrix of outliers.

The rest of this paper is structured as follows. In Section
\ref{secRelatW}, we introduce the mathematical model of the data and
discuss related works. In Section \ref{secSOR}, we describe the
proposed CoR-UTV method in detail. In Section \ref{secRobustPCA}, we
develop an algorithm for robust PCA using CoR-UTV. In Section
\ref{secNumExp}, we present and discuss simulation results and
conclusions are given in Section \ref{secCon}.

\section{Mathematical Model and Related Works}
\label{secRelatW}
Given a matrix ${\bf A} \in \mathbb R^{m \times
n}$, where $m \ge n$, with numerical rank $k$, its SVD
\cite{GolubVanLoan96} is defined as:
\begin{equation}
\begin{aligned}
{\bf A} = & {\bf U}_\text{A}{\bf \Sigma}_\text{A}{\bf V}_\text{A}^T
= & \underbrace{\begin{bmatrix} {{\bf U}_k \quad {\bf U}_0}
\end{bmatrix}}_{{\bf U}_\text{A} \in \mathbb R^{m \times n}}
  \underbrace{\begin{bmatrix}
       {\bf \Sigma}_k & 0  \\
       0 & {\bf \Sigma}_0
  \end{bmatrix}}_{{\bf \Sigma}_\text{A} \in \mathbb R^{n \times n}}
  \underbrace{\begin{bmatrix}{{\bf V}_k \quad {\bf V}_0} \end{bmatrix}^T}_{{\bf V}_\text{A}^T \in \mathbb R^{n \times n}},
\label{eqSVD}
\end{aligned}
\end{equation}
where ${\bf U}_k \in \mathbb R^{m \times k}$, ${\bf U}_0 \in \mathbb
R^{m \times n-k}$, ${\bf V}_k \in \mathbb R^{n \times k}$ and ${\bf
V}_0 \in \mathbb R^{n \times n-k}$ have orthonormal columns, ${\bf
\Sigma}_k \in \mathbb R^{k \times k}$ and ${\bf \Sigma}_0 \in
\mathbb R^{n-k \times n-k}$ are diagonal matrices containing the
singular values, i.e., ${\bf \Sigma}_k=\text{diag}(\sigma_1, ...,
\sigma_k)$ and ${\bf \Sigma}_0 =\text{diag}(\sigma_{k+1}, ...,
\sigma_n)$. $\bf A$ can be written as ${\bf A} = {\bf A}_k+{\bf
A}_0$, where ${\bf A}_k = {\bf U}_k{\bf \Sigma}_k{\bf V}_k^T$, and
${\bf A}_0 = {\bf U}_0{\bf \Sigma}_0{\bf V}_0^T$. The SVD constructs
the optimal rank-$k$ approximation ${\bf A}_k$ to ${\bf A}$,
\cite{GolubVanLoan96} i.e., \vspace*{-1.2mm}
\begin{equation}
\begin{aligned}
&\ \|{\bf A} - {\bf A}_k\|_2 = \sigma_{k+1},  \\
&\ {\|{\bf A} - {\bf A}_k\|_F} = \sqrt{\sigma_{k+1}^2 +...+ \sigma_{n}^2},
\end{aligned}\label{equ9}
\end{equation}
where ${\|{\cdot}\|_2}$ and ${\|{\cdot}\|_F}$ denote the spectral  norm and the Frobenius norm, respectively. In this paper we focus on the matrix $\bf A$ defined.%\looseness-1

The SVD is highly accurate for computing singular subspaces and
singular values. However, its computation is costly for large data
sets. Moreover, standard techniques for its computation are
challenging to parallelize in order to take advantage of modern
processors \cite{HMT2009, Gu2015}. An economical version of the SVD
is the partial SVD based on Krylov subspace methods, such as the
Lanczos and Arnoldi algorithms, which constructs an approximate SVD
of an input matrix, for instance $\bf A$, at a cost $O(mnk)$.
However, the partial SVD suffers from two drawbacks: (i) it is
numerically unstable \cite{CalvettiRS94,GolubVanLoan96}, and (ii) it
does not lend itself to parallel implementations \cite{HMT2009,
Gu2015}, which makes it unsuitable for modern architectures. Other
approaches for low-rank matrix approximations include the RRQR
\cite{Chan87} and the UTV decompositions \cite{Stewart98}. Even
though the QR with column pivoting (QRCP) and UTV decompositions
provide highly accurate approximations to $\bf A$, they suffer from
two drawbacks: (i) they are costly, i.e., $O(mn^2)$, and (ii)
methods for their computation are challenging to parallelize and
hence they cannot exploit modern computational platforms
\cite{HMT2009,Gu2015}.

Recently developed algorithms for low-rank approximations based on
randomization  \cite{FriezeKVS04, Rokhlin09, HMT2009,
Gu2015,TrYUC17} have attracted significant attention. The randomized algorithms project a large input matrix onto a lower dimensional space using a random matrix, and apply deterministic methods on the smaller matrix to give an approximation of the matrix. Hence (i) they are computationally efficient, and (ii) lend themselves to parallel implementation. Halko et al. \cite{HMT2009} proposed \textit{randomized SVD} (R-SVD) in which a smaller matrix is formed by linear combinations of columns of the given matrix. The low-rank approximation is then given through the SVD of a reduced-size matrix.  Gu \cite{Gu2015} applied a slightly modified version of the R-SVD algorithm to improve subspace iteration methods, and presents a new error analysis. Another algorithm proposed in \cite[Section 5.5]{HMT2009}, which we call two-sided randomized SVD (TSR-SVD), is a \textit{single-pass} method, i.e., it required only one pass through data. It captures most attributes of the data by means of forming the smaller matrix through linear combinations of both rows and columns of the given matrix, and then applies the SVD for further computations. The work in \cite{MFKDeTSP18} proposed a randomized algorithm termed subspace-orbit randomized SVD (SOR-SVD). SOR-SVD alternately projects the matrix onto its column and row space. The matrix is then transformed into a lower dimensional space, and a truncated SVD follows in order to construct an approximation.

TSR-SVD gives poor approximation compared to the optimal SVD due to the single-pass strategy. SOR-SVD has shown better performance than TSR-SVD, however both methods apply the SVD on the reduced-size matrix. This computation may be burdensome in terms of communication cost \cite{DemmGHL12} for large matrices. In this work, we develop a randomized algorithm for low-rank approximation that with comparable flops (i) outperforms the TSR-SVD in terms of accuracy, and (ii) can utilize advanced computer architectures better than TSR-SVD as well as SOR-SVD.

\section{Compressed Randomized UTV Decompositions}
\label{secSOR}

In this section, we present a randomized rank-revealing
decomposition algorithm termed compressed randomized UTV (CoR-UTV)
decomposition \cite{corutv}, which computes a low-rank approximation
of a given matrix. We focus on the matrix $\bf A$ with $m \ge n$,
where CoR-UTV, in the form of \eqref{eq_contri}, produces an upper
triangular middle matrix $\bf T$. The modifications required for a
CoR-UTV for the case $m < n$ that produces a lower triangular middle
matrix $\bf T$ is straightforward.
%the URV and ULV decompositions see \cite{Hansen98, FierroHan97} and
%the references therein. We also present a version of CoR-UTV with
%power iteration, which improves the performance of the algorithm at
%an extra computational cost.

\subsection{Proposed CoR-UTV Decompositions}

Given the matrix ${\bf A}$ and an integer $k\le \ell<
\text{min}\{m,n\}$, the basic version of CoR-UTV is computed as
follows: using a random number generator, we form a matrix ${\bf
\Psi} \in \mathbb R^{n \times \ell}$ with entries drawn independent,
identically distributed (i.i.d.) from the standard Gaussian
distribution. We then compute the matrix product:
\begin{equation}
{\bf C}_1 = {\bf A}{\bf \Psi},
\label{eq_C1_1st}
\end{equation}
where ${\bf C}_1 \in \mathbb R^{m \times \ell}$ is, in fact, a
projection onto the subspace spanned by columns of ${\bf A}$. Having
${\bf C}_1$, we form ${\bf C}_2 \in \mathbb R^{n \times \ell}$:
\begin{equation}
{\bf C}_2 = {\bf A}^T{\bf C}_1,
\label{eqAC1}
\end{equation}
where ${\bf C}_2$ is, in fact, a projection onto the subspace
spanned by rows of ${\bf A}$. Using a QR decomposition, we factor
${\bf C}_1$ and ${\bf C}_2$ such that:
\begin{equation}
{\bf C}_1 = {\bf Q}_1{\bf R}_1,  \quad \text{and} \quad {\bf C}_2 =
{\bf Q}_2{\bf R}_2,
\end{equation}
where ${\bf Q}_1$ and ${\bf Q}_2$ are approximate bases for
$\mathcal{R}({\bf A})$ and $\mathcal{R}({\bf A}^T)$, respectively.
We now compress $\bf A$ by left and right multiplications by
the orthonormal bases obtained, forming the matrix ${\bf D} \in
\mathbb R^{\ell \times \ell}$:
\begin{equation}
{\bf D}={\bf Q}_1^T{\bf A}{\bf Q}_2,
\label{eqM}
\end{equation}
We then compute a QRCP of ${\bf D}$:
\begin{equation}
{\bf D} = \widetilde{\bf Q}\widetilde{\bf R}\widetilde{\bf P}^T.
\label{eq_Drrqr}
\end{equation}
The CoR-UTV-based low-rank approximation of $\bf A$ is given by
\begin{equation}
\hat{\bf A}_\text{CoR}= {\bf UTV}^T,
\label{eq_T_basic}
\end{equation}
where ${\bf U}={\bf Q}_1 \widetilde{\bf Q} \in \mathbb R^{m \times
\ell}$ and ${\bf V}={\bf Q}_2  \widetilde{\bf P} \in \mathbb R^{n
\times \ell}$ construct approximations to the $\ell$ leading left
and right singular vectors of $\bf A$, respectively, and ${\bf
T}=\widetilde{\bf R}\in \mathbb R^{\ell \times \ell}$ is upper
triangular with diagonals approximating the first $\ell$ singular
values of $\bf A$.

CoR-UTV requires three passes through data, for a matrix stored
externally, but it can be altered to revisit the data only once. To
this end, the compressed matrix $\bf D$ \eqref{eqM} can be
approximated as follows: both sides of the currently unknown ${\bf
D}={\bf Q}_1^T{\bf A}{\bf Q}_2$ are postmultiplied by ${\bf
Q}_2^T{\bf \Psi}$. Having defined ${\bf A}\approx {\bf A}{\bf
Q}_2{\bf Q}_2^T$ and ${\bf C}_1 = {\bf A}{\bf \Psi}$, then $ {\bf
D}_\text{approx} = {\bf Q}_1^T{\bf C}_1({\bf Q}_2^T{\bf
\Psi})^\dagger$.

CoR-UTV is accurate for matrices whose singular values display some
decay, however in applications where the data matrix has a slowly
decaying singular spectrum, it may produce a poor approximation
compared to the SVD. Thus, we incorporate $q$ steps of a power
iteration \cite{Rokhlin09,HMT2009} to improve the accuracy of the
algorithm in these circumstances. Given the matrix ${\bf A}$, and
integers $k\le \ell< n$ and $q$, the resulting algorithm is
described in Alg. \ref{Alg3}.
\begin{algorithm}
\caption{CoR-UTV with Power Method}
\renewcommand{\algorithmicrequire}{\textbf{Input:}}
\begin{algorithmic}[1]
\REQUIRE ~~ % ：Input
 Matrix $\ {\bf A} \in \mathbb R^{m \times n}$,
integers $k$, $\ell$ and $q$.
\renewcommand{\algorithmicrequire}{\textbf{Output:}}
\REQUIRE ~~ A rank-$\ell$ approximation.
  \STATE Draw a standard Gaussian matrix ${\bf C}_2 \in \mathbb R^{n \times \ell}$;
  \FOR{$i=$ 1: $q+1$}
   \STATE Compute ${\bf C}_1 = {\bf A}{\bf C}_2$; \\
   \STATE Compute ${\bf C}_2 = {\bf A}^T{\bf C}_1$;
  \ENDFOR \\
  \STATE Compute QR decompositions ${\bf C}_1 = {\bf Q}_1{\bf R}_1$,
  ${\bf C}_2 = {\bf Q}_2{\bf R}_2$;
  \STATE Compute ${\bf D}={\bf Q}_1^T{\bf A}{\bf Q}_2$ or ${\bf D}_\text{approx} = {\bf Q}_1^T{\bf C}_1({\bf Q}_2^T{\bf C}_2)^
  \dagger $;
  \STATE Compute a QRCP ${\bf D} = \widetilde{\bf Q}\widetilde{\bf R}\widetilde{\bf P}^T$ or ${\bf D}_\text{approx} = \widetilde{\bf Q}\widetilde{\bf R}\widetilde{\bf P}^T$;
  \STATE Form the CoR-UTV-based low-rank approximation of $\bf A$:
  $\hat{\bf A}_\text{CoR}= {\bf UTV}^T$; ${\bf U}={\bf Q}_1 \widetilde{\bf Q},
  {\bf T}=\widetilde{\bf R}$,${\bf V}={\bf Q}_2\widetilde{\bf P}^T$.
\end{algorithmic}\label{Alg3}
\end{algorithm}

\subsection{Computational Complexity}

The cost of an algorithm involves both arithmetic, i.e., the number
of flops, and communication, i.e., data movement either between
different levels of a memory hierarchy or between processors
\cite{DemmGHL12}. On multicore and accelerator-based computers, for
a data matrix stored externally, the communication cost becomes
substantially more expensive compared to the arithmetic
\cite{DemmGHL12, Dongarra17}. The randomized algorithms operate on a
compressed version of the data matrix rather than a matrix itself
and therefore can be organized to exploit modern computational
environments better than their classical counterparts.

To decompose $\bf A$, the simple version of CoR-UTV incurs the
following costs: Step 1 costs $n\ell$, Step 2 costs $2mn\ell$, Step
3 costs $2mn\ell$, Step 4 costs $2m\ell^2 + 2n\ell^2$, Step 5 costs
$m\ell^2+2mn\ell$ (if $\bf D$ is approximated by ${\bf
D}_\text{approx}$, this step costs $2m\ell^2 + 2n\ell^2 +3\ell^3$),
Step 6 costs $8/3\ell^3$, Step 7 costs $2m\ell^2 + 2n\ell$. The
dominant cost of Steps 1-7 occurs when multiplying $\bf A$ and ${\bf
A}^T$ with the corresponding matrices. Thus
\begin{equation}
C_\text{CoR-UTV} = O(mn\ell).
\label{equCost1}
\end{equation}

The sample size parameter $\ell$ is typically close to the minimal
rank $k$. The simple form of CoR-UTV requires either three or two
passes (when $\bf D$ is approximated by ${\bf D}_\text{approx}$)
through data to factor $\bf A$. When the power method is
incorporated, CoR-UTV requires either $(2q+3)$ or $(2q+2)$ passes
(when $\bf D$ is approximated by ${\bf D}_\text{approx}$) over the
data with arithmetic costs of $(2q+3)C_\text{CoR-UTV}$ or
$(2q+2)C_\text{CoR-UTV}$, respectively.

In addition to matrix-matrix multiplications and QR decompositions,
CoR-UTV performs one QRCP on an $\ell \times \ell$ matrix, however
TSR-SVD and SOR-SVD perform an SVD on the $\ell \times \ell$ matrix.
The SVD is more expensive than QRCP and, furthermore, recently
developed QRCP algorithms based on randomization can perform the
factorization with minimum communication costs \cite{DemGGX15,
DuerschGu2017,MartinssonHQRRP2017}, while standard techniques to
compute an SVD are challenging for parallelization
\cite{HMT2009,Gu2015}. Hence for large matrices to be factored on
high performance computing architectures, where the compressed $\ell
\times \ell$ matrix does not fit into fast memory, the execution
time to compute CoR-UTV can be substantially less than those of
TSR-SVD and SOR-SVD. This is an advantage of CoR-UTV over TSR-SVD
and SOR-SVD.

\section{Robust PCA with CoR-UTV}
\label{secRobustPCA}

This section describes how to solve the robust PCA problem using the
proposed CoR-UTV method. Robust PCA \cite{CSPW2009, CLMW2009}
represents an input low-rank matrix ${\bf M} \in \mathbb R^{m \times
n}$ whose fraction of entries being corrupted, as a linear
superposition of a low-rank matrix ${\bf L}$ and a sparse matrix of
outliers ${\bf S}$ such as ${\bf M=L+S}$, by solving the following
convex program:
\begin{equation}
\begin{aligned}
&{\text{minimize}_{\bf(L, S)}} \ {\|{\bf L}\|_* + \lambda\|{\bf S}\|_1} \\
&{\text{subject to}} \ {\bf M} = {\bf L} + {\bf S},
\end{aligned}\label{equV1}
\end{equation}
where ${\|\mbox{\bf B}\|_*}  \triangleq \sum_i\sigma_i (\mbox{\bf
B}) $ is the nuclear norm of any matrix $\mbox{\bf B}$,
${\|\mbox{\bf B}\|_1} \triangleq \sum_{ij} |\mbox{\bf B}_{ij}|$ is
the $\ell_{1}$-norm of $\mbox{\bf B}$, and $\lambda>0$ is a tuning
parameter. One efficient method to solve \eqref{equV1} is the method
of augmented Lagrange multipliers (ALM) \cite{Bertsekas1982}. The
ALM method yields the optimal solution, however its bottleneck is
computing the costly SVD at each iteration to approximate the
low-rank component $\bf L$ of $\bf M$ \cite{CLMW2009, LLS2011}. To
address this issue and to speed up the convergence of the ALM
method, the work in \cite{LLS2011} proposes a few techniques
including predicting the principal singular space dimension, a
continuation technique \cite{Toh2010}, and a truncated SVD by using
PROPACK package \cite{Larsen98}. The modified algorithm
\cite{LLS2011} substantially improves the convergence speed, however
the truncated SVD \cite{Larsen98} employed uses the Lanczos
algorithm that (i) is unstable, and (ii) due to the limited data
reuse in its operations, has very poor performance on modern
architectures \cite{CalvettiRS94, GolubVanLoan96,HMT2009,Gu2015}.

To address this issue, by considering the original objective
function proposed in \cite{CSPW2009, CLMW2009, LLS2011}, we apply
CoR-UTV as a surrogate to the truncated SVD to solve the robust PCA
problem. We adopt the continuation technique \cite{Toh2010,
LLS2011}, which increases $\mu$ in each iteration. The proposed
\texttt{ALM-CoRUTV} method is given in Alg. \ref{TableALM-CoRUTV}.

\begin{algorithm}
\caption{Robust PCA solved by \texttt{ALM-CoRUTV}}
\renewcommand{\algorithmicrequire}{\textbf{Input:}}
\begin{algorithmic}[1]
\REQUIRE ~~ % ：Input
 Matrix ${\bf M}, \lambda, \mu, {\bf Y}_0 = {\bf S}_0 = 0, j=0$.
\renewcommand{\algorithmicrequire}{\textbf{Output:}}
\REQUIRE ~~ Low-rank plus sparse matrix.
\WHILE {the algorithm does not converge}
       \STATE Compute ${\bf L}_{j+1} = \mathcal{C}_{\mu_j^{-1}}
        ({\bf M} - {\bf S}_j +\mu_j^{-1} {\bf Y}_j)$;
        \STATE Compute ${\bf S}_{j+1} = \mathcal{S}_{\lambda\mu_j^{-1}}
        ({\bf M} - {\bf L}_{j+1} +\mu_j^{-1} {\bf Y})$;
        \STATE Compute ${\bf Y}_{j+1} = {\bf Y}_j +\mu_j({\bf M} - {\bf L}_{j+1}
        - {\bf S}_{j+1})$;
        \STATE Update $\mu_{j+1} = \text{max}(\rho\mu_j, {\bar \mu})$;
\ENDWHILE
\RETURN $\bf L^*$ and $\bf S^*$.
\end{algorithmic}\label{TableALM-CoRUTV}
\end{algorithm}

In Alg. \ref{TableALM-CoRUTV}, for any matrix $\bf B$ with a CoR-UTV
decomposition described in Section \ref{secSOR}, $\mathcal{C}_\delta
({\bf B})$ refers to a CoR-UTV thresholding operator defined as:
\begin{equation}
\mathcal{C}_\delta({\bf B})={\bf U}(:,1:r){\bf T}(1:r,:){\bf V}^T,
\end{equation}
where $r$ is the number of diagonals of $\bf T$ greater than
$\delta$, $\mathcal{D}_\delta ({\bf B})$ refers to a singular value
thresholding operator defined as $\mathcal{D}_\delta ({\bf B}) =
{\bf U}_\text{B}\mathcal{S}_\delta ({\bf \Sigma}_\text{B}){\bf
V}_\text{B}^T$, where $\mathcal{S}_\delta (x) =
{\text{sgn}(x)\text{max}}(|x| - \delta, 0)$ is a shrinkage operator
\cite{Hale2008}, $\lambda$, $\mu_0$, ${\bar \mu}$, $\rho$, ${\bf
Y}_0$, and ${\bf S}_0$ are initial values. The main operation of
\texttt{ALM-CoRUTV} is computing CoR-UTV in each iteration, which is
efficient in terms of flops, $O(mnk)$, and can
be computed with minimum communication costs.%; see subsection
%\ref{secComComplex}. In subsection \ref{subrpca}, we experimentally
%verify that \texttt{ALM-CoRUTV} converges to the exact optimal
%solution.

\section{Numerical Experiments}
\label{secNumExp}

In this section, we present simulations that evaluate the
performance of CoR-UTV for approximating a low-rank input matrix. We
show that CoR-UTV provides highly accurate singular values and
low-rank approximations, and compare CoR-UTV against competing
algorithms from the literature. We also employ CoR-UTV for
solving the robust PCA problem.% The experiments were run in MATLAB
%on a desktop PC with a 3 GHz intel Core i5-4430 processor and 8 GB
%of memory.

\subsection{Rank-Revealing Property \& Singular Values Estimation}

We first show that CoR-UTV (i) is rank revealer, i.e., the gap in
the singular value spectrum of the matrix is revealed, and (ii)
provides highly accurate singular values. For the randomized
algorithms considered, namely CoR-UTV, TSR-SVD, and SOR-SVD, the
results presented are averaged over 20 trials. Each trial was run
with the same input matrix with an independent draw of the test
matrix. Due to space constraints, we only consider one class of
low-rank matrices, and for simplicity we focus on a square matrix.

We construct a noisy rank-$k$ matrix ${\bf A}$ of order $10^3$
generated as ${\bf A} ={\bf U\Sigma V}^T + 0.1\sigma_k{\bf E}$,
where ${\bf U}$ and ${\bf V}$ are random orthonormal matrices, ${\bf
\Sigma}$ is diagonal containing the singular values $\sigma_i$s that
decrease linearly from $1$ to $10^{-9}$,
$\sigma_{k+1}=...=\sigma_{10^3}=0$, and ${\bf E}$ is a normalized
Gaussian matrix. We set $k=20$.

We compare the singular values of the matrix computed by CoR-UTV
against those of competing methods such as the SVD
\cite{GolubVanLoan96}, QRCP \cite{Chan87}, UTV \cite{FierroHanHan99}
and TSR-SVD \cite{HMT2009}. For CoR-UTV and TSR-SVD, we arbitrarily
set the sample size parameter to $\ell=2k$. Both algorithms require
the same number of passes over $\bf A$, either two or $2q+2$ when
the power method is used, to perform a
factorization. %To compute a UTV decomposition, we implement the
%\texttt{lurv} function from \cite{FierroHanHan99}.

\begin{figure}[t]
\begin{center}
\def\epsfsize#1#2{1\columnwidth}
\epsfbox{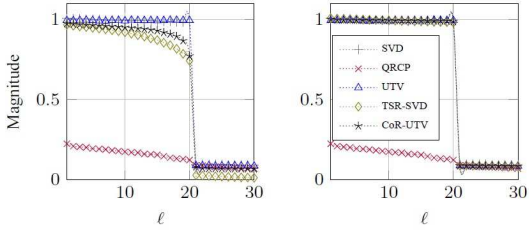}
\captionsetup{justification=centering,font=scriptsize}
\vspace{-0.95em}\caption{{Comparison of singular values. $q=0$ (left), and $q=2$ (right).}}%: CoR-UTV
%strongly reveals the gap in the singular values, as do the SVD, UTV
%and TSR-SVD, while QRCP only suggests the gap. CoR-UTV outperforms
%TSR-SVD in estimating leading and trailing singular values. Right
%($q=2$): With two steps of a power method CoR-UTV delivers singular
%values as accurate as the optimal SVD. QRCP considerably
%underestimates the leading singular values of the matrix.}}
\label{fig:SV_M1_2}       % Give a unique label
\end{center}
\end{figure}

The results are shown in Fig. \ref{fig:SV_M1_2}. It is observed that
(i) CoR-UTV strongly reveals the numerical rank $k$, (ii) with no
power iterations ($q=0$), CoR-UTV provides very good approximations
to singular values and outperforms TSR-SVD in approximating both
leading and trailing singular values, (iii) with $q=2$, CoR-UTV
delivers singular values as accurate as the optimal SVD, (iv) QRCP
only suggests the gap in the singular spectrum, and gives a fuzzy
approximation to singular values of the matrix.
%\vspace{-.3cm}

\subsection{Low-Rank Approximation}

We now compare the low-rank approximation constructed by our method
against those of the SVD, QRCP, TSR-SVD, and SOR-SVD
\cite{MFKDeTSP18}. We construct a rank-$k$ approximation ${\hat{\bf
A}}_\text{out}$ to ${\bf A}$ by varying the sample size parameter
$\ell$ with the rank fixed, and calculate the error:
\begin{equation}
e_k = \|{\bf A} - \hat{\bf A}_{\text{out}}\|_F.
\end{equation}

\begin{figure}[t]
\begin{center}
\def\epsfsize#1#2{1\columnwidth}
\epsfbox{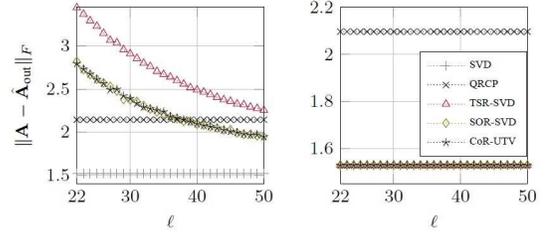} \caption{Comparison of low-rank approximation
errors. $q=0$ (left), and $q=2$ (right).}
\label{fig:LR_M1II}       % Give a unique label
\end{center}
\end{figure}

The results are shown in Fig. \ref{fig:LR_M1II}. It is observed that
(i) when $q=0$, CoR-UTV and SOR-SVD show similar performances, while
TSR-SVD shows the worst performance, (ii) when $q=2$, the errors
resulting from CoR-UTV show no loss of accuracy compared to the
optimal SVD. In this case, QRCP has the poorest performance.

\subsection{Robust Principal Component Analysis}
\label{subrpca}

Here, we examine the efficiency and efficacy of \texttt{ALM-CoRUTV}
in Alg. \ref{TableALM-CoRUTV} for recovering the low-rank and sparse
components of data. We compare the results obtained with those of
the efficient inexact ALM method by \cite{LLS2011}, called
\texttt{InexactALM} hereafter.

%We form a rank-$k$ matrix $\bf M =L+S$ as a linear combination of a
%low-rank matrix ${\bf L} \in \mathbb R^{n \times n}$ and a sparse
%error matrix  ${\bf S}\in \mathbb R^{n \times n}$.
Robust PCA represents an input low-rank matrix ${\bf M} \in \mathbb
R^{m \times n}$, whose a fraction of entries being corrupted, as a
linear superposition of a low-rank matrix ${\bf L}$ and a sparse
matrix of outliers ${\bf S}$ such as ${\bf M=L+S}$, by solving the
following convex program:
\begin{equation}
\begin{aligned}
&{\text{minimize}_{\bf(L, S)}} \ {\|{\bf L}\|_* + \lambda\|{\bf S}\|_1} \\
&{\text{subject to}} \ {\bf M} = {\bf L} + {\bf S},
\end{aligned}\label{equV1}
\end{equation}
where ${\|\mbox{\bf B}\|_*}  \triangleq \sum_i\sigma_i (\mbox{\bf
B}) $ is the nuclear norm of any matrix $\mbox{\bf B}$,
${\|\mbox{\bf B}\|_1} \triangleq \sum_{ij} |\mbox{\bf B}_{ij}|$ is
the $\ell_{1}$-norm of $\mbox{\bf B}$, and $\lambda>0$ is a tuning
parameter
\cite{rstap2012,sastap2012,locsme2014,saalt2014,als2015,dce2015,damdc2016,okspme2016,bfpeg2011,vfap2012,memd2016},
\cite{spa2008,gbd2013,tds2012,lsmimo2015,mfsic2011,mbthp2014,mbdf2013,mfdf2012,armo2013,sicdma2011,mbsic2011,bfidd2016},
\cite{rmbthp2014,did2014,jpais2012,badstbc2016,wlbd2017,mmimo2013,bbprec2017,1bitidd2018,baplnc2018}.
The matrix ${\bf L}$ is generated as ${\bf L}={\bf U}{\bf V}^T$,
where ${\bf U}$, ${\bf V} \in \mathbb R^{n \times k}$ have standard
Gaussian distributed entries. The error matrix ${\bf S}$ has $s$
non-zero entries independently drawn from the set $\lbrace$-80,
80$\rbrace$. We apply the \texttt{ALM-CoRUTV} and
\texttt{InexactALM} algorithms to $\bf M$ to recover ${\bf L}$ and
${\bf S}$. The numerical results are summarized in Table
\ref{TableCoR1}, where the rank of $\bf L$ $r({\bf L})=0.05\times n$
and $s = \|{\bf S}\|_0=0.05\times n^2$.

In our experiments, we adopt the initial values suggested in
\cite{LLS2011}. The algorithms are terminated when $ {\|{\bf M}-{\bf
L}^{out}-{\bf S}^{out}\|_F}< 10^{-5}{\|{\bf M}\|_F}$ is satisfied,
where $({\bf L}^{out}, {\bf S}^{out})$ is the pair of output of
either algorithm. In the Table, $Time(s)$ refers to the runtime in
seconds, $Iter.$ refers to the number of iterations, and
$\zeta={\|{\bf M}-{\bf L}^{out}-{\bf S}^{out}\|_F}/{{\|{\bf
M}\|_F}}$ refers to the relative error.

\begin{table}[!htb]
\vspace{-0.75em}\caption{Numerical results for synthetic matrix
recovery.} \vspace{-0.5cm}
\begin{tabular}
{p{0.3cm} p{0.3cm} p{0.4cm} p{1.6cm} p{0.4cm} p{0.53cm} p{0.7cm} p{0.3cm} p{0.2cm}}
\noindent\rule{8.4cm}{0.4pt}\\
$n$ & $r(\bf L)$ & $\|{\bf S}\|_0$ & Methods &
$r({\bf L}^*)$ & $\|{\bf S}^*\|_0$ & Time(s) & Iter.& $\zeta$ \\
\noindent\rule{8.4cm}{0.4pt}\\
1000& 50 & 5e4 &
\begin{tabular}{|c p{0.2cm} p{0.7cm} p{0.4cm} p{0.1cm} p{0.9cm}}
\texttt{InexactALM} & 50& 5e4 & 4.1 & 12 & 2.1e-6  \\
\texttt{ALM-CoRUTV} & 50& 5e4 & 0.6 & 12 & 9.6e-6 \\
\end{tabular} \\
2000& 100 & 2e5 &
\begin{tabular}{|c p{0.2cm} p{0.7cm} p{0.4cm} p{0.1cm} p{0.9cm}}
\texttt{InexactALM} & 100& 2e5 & 27.4 & 12 & 2.7e-6  \\
\texttt{ALM-CoRUTV} & 100& 2e5 & 3.7  & 12 & 8.3e-6 \\
\end{tabular} \\
3000& 150 & 45e4 &
\begin{tabular}{|c p{0.2cm} p{0.7cm} p{0.4cm} p{0.1cm} p{0.9cm}}
\texttt{InexactALM} & 150& 45e4 & 75.6 & 12 & 3.1e-6  \\
\texttt{ALM-CoRUTV} & 150& 45e4 & 9.4  & 12 & 8.7e-6 \\
\end{tabular} \\
\noindent\rule{8.4cm}{0.4pt}
\end{tabular}
\label{TableCoR1}
\end{table}
%\vspace{-.5cm}
CoR-UTV requires a prespecified rank $\ell$ to perform the
factorization. Thus, we set $\ell=2k$, as a random start, and $q=1$.
The results in Table \ref{TableCoR1} show that \texttt{ALM-CoRUTV}
detects the exact rank $k$ of the input matrix, provides the exact
optimal solution, and outperforms \texttt{InexactALM} in terms of
runtime.

\section{Conclusions}
\label{secCon}

In this paper, we have presented CoR-UTV for computing low-rank
approximations of an input matrix. Simulations show that CoR-UTV
reveals the numerical rank more sharply than QRCP, and provides
results as good as those of the optimal SVD. CoR-UTV is more
efficient than SVD, QRCP, UTV, TSR-SVD and SOR-SVD in terms of cost.
CoR-UTV can better exploit advanced computational platforms by
leveraging higher levels of parallelism than all compared
algorithms. We have also applied CoR-UTV to solve the robust PCA
problem via the ALM method. Simulations show that
\texttt{ALM-CoRUTV} outperforms efficiently implemented
\texttt{InexactALM}.

%\appendices
%\section{The spectral-norm error comparison for rank-$k$ approximation}
%This appendix provides numerical results on the performance of the algorithms discussed in subsection \ref{subs_RankKapp} for the rank-$k$ approximation of Matrix 1 and Matrix 2 in terms of the spectral-norm error.
%
%\begin{figure}[t]
%\begin{center}
%\input{graph/res_LowRank_L2NMat1_I}
%\caption{Comparison of the spectral-norm rank-$k$ approximation error for  \texttt{NoisyLowRank-I}. Left: No power method, $q=0$. Right: $q=2$.}
%\label{fig:LR_L2M1_I}       % Give a unique label
%\end{center}
%\end{figure}
%
%\begin{figure}[t]
%\begin{center}
%\input{graph/res_LowRank_L2NMat1_II}
%\caption{Comparison of the spectral-norm rank-$k$ approximation error for  \texttt{NoisyLowRank-II}. Left: No power method, $q=0$. Right: $q=2$.}
%\label{fig:LR_L2M1_II}       % Give a unique label
%\end{center}
%\end{figure}
%
%\begin{figure}[t]
%\begin{center}
%\input{graph/res_LowRank_L2NMat2}
%\caption{Comparison of the spectral-norm rank-$k$ approximation error for Matrix 2. Left: No power method, $q=0$. Right: $q=2$.}
%\label{fig:LR_L2M2}       % Give a unique label
%\end{center}
%\end{figure}

{\small
\bibliographystyle{IEEEtran}
\bibliography{mybibfile}}

\begin{thebibliography}{10}
\providecommand{\url}[1]{#1}
\csname url@samestyle\endcsname
\providecommand{\newblock}{\relax}
\providecommand{\bibinfo}[2]{#2}
\providecommand{\BIBentrySTDinterwordspacing}{\spaceskip=0pt\relax}
\providecommand{\BIBentryALTinterwordstretchfactor}{4}
\providecommand{\BIBentryALTinterwordspacing}{\spaceskip=\fontdimen2\font plus
\BIBentryALTinterwordstretchfactor\fontdimen3\font minus
  \fontdimen4\font\relax}
\providecommand{\BIBforeignlanguage}[2]{{%
\expandafter\ifx\csname l@#1\endcsname\relax
\typeout{** WARNING: IEEEtran.bst: No hyphenation pattern has been}%
\typeout{** loaded for the language `#1'. Using the pattern for}%
\typeout{** the default language instead.}%
\else
\language=\csname l@#1\endcsname
\fi
#2}}
\providecommand{\BIBdecl}{\relax}
\BIBdecl

\bibitem{8425659}
T.~Bouwmans, S.~Javed, H.~Zhang, Z.~Lin, and R.~Otazo, ``On the applications of
  robust pca in image and video processing,'' \emph{Proceedings of the IEEE},
  vol. 106, no.~8, pp. 1427--1457, Aug 2018.

\bibitem{KaDeDSP17}
M.~F. Kaloorazi and R.~C. de~Lamare, ``{Low-Rank and Sparse Matrix Recovery
  Based on a Randomized Rank-revealing Decomposition},'' in \emph{22nd Intl
  Conf. on DSP 2017, UK}, Aug 2017.

\bibitem{FazelPST13}
M.~Fazel, T.~K. Pong, D.~Sun, and P.~Tseng, ``{Hankel Matrix Rank Minimization
  with Applications to System Identification and Realization},'' \emph{SIAM. J.
  Matrix Anal. \& Appl.}, vol.~34, no.~3, pp. 946--–977, Apr 2013.

\bibitem{KaDeICASSP17}
M.~F. Kaloorazi and R.~C. de~Lamare, ``{Anomaly Detection in IP Networks Based
  on Randomized Subspace Methods},'' in \emph{ICASSP}, 2017.

\bibitem{NHussainP18}
A.~M.~J. Niyaz~Hussain and M.~Priscilla, ``{A Survey on Various Kinds of
  Anomalies Detection Techniques in the Mobile Adhoc Network Environment},''
  \emph{Int J S Res CSE \& IT.}, vol.~3, no.~3, pp. 1538--1541, 2018.

\bibitem{Chandrasekaran12}
V.~Chandrasekaran, P.~Parrilo, and A.~Willsky, ``{Latent variable graphical
  model selection via convex optimization},'' \emph{The Ann. of Stat.},
  vol.~40, no.~4, pp. 1935–--1967.

\bibitem{RahmaniAtiaCoP17}
M.~Rahmani and G.~Atia, ``{Coherence Pursuit: Fast, Simple, and Robust
  Principal Component Analysis},'' \emph{IEEE Trans. Signal Process.}, vol.~65,
  no.~23, pp. 6260--6275, Dec 2017.

\bibitem{Oh2017}
T.-H. Oh, Y.~Matsushita, Y.-W. Tai, and I.~S. Kweon, ``Fast randomized singular
  value thresholding for low-rank optimization,'' \emph{IEEE Trans. Pattern
  Anal. Mach. Intell.}, vol.~40, no.~2, pp. 376--391, Mar. 2017.

\bibitem{ifir2005}
R.~C. de~Lamare and R.~Sampaio-Neto, ``Reduced-rank interference suppression
  for ds-cdma based on interpolated fir filters,'' \emph{IEEE Communications
  Letters}, vol.~9, no.~3, pp. 213--215, March 2005.

\bibitem{intadap2005}
------, ``Adaptive reduced-rank mmse filtering with interpolated fir filters
  and adaptive interpolators,'' \emph{IEEE Signal Processing Letters}, vol.~12,
  no.~3, pp. 177--180, March 2005.

\bibitem{jio2007}
------, ``Reduced-rank adaptive filtering based on joint iterative optimization
  of adaptive filters,'' \emph{IEEE Signal Processing Letters}, vol.~14,
  no.~12, pp. 980--983, Dec 2007.

\bibitem{jiomvdr2008}
R.~C. de~Lamare, ``Adaptive reduced-rank lcmv beamforming algorithms based on
  joint iterative optimisation of filters,'' \emph{Electronics Letters},
  vol.~44, no.~9, pp. 565--566, April 2008.

\bibitem{mwfccm2008}
R.~C. de~Lamare, M.~Haardt, and R.~Sampaio-Neto, ``Blind adaptive constrained
  reduced-rank parameter estimation based on constant modulus design for cdma
  interference suppression,'' \emph{IEEE Transactions on Signal Processing},
  vol.~56, no.~6, pp. 2470--2482, June 2008.

\bibitem{DeSa2009}
R.~de~Lamare and R.~Sampaio-Neto, ``{Adaptive Reduced-Rank Processing Based on
  Joint and Iterative Interpolation, Decimation, and Filtering},'' \emph{IEEE
  Trans. Signal Process.}, vol.~57, no.~7, pp. 2503--2514.

\bibitem{jiols2010}
R.~C. de~Lamare and R.~Sampaio-Neto, ``Reduced-rank space-time adaptive
  interference suppression with joint iterative least squares algorithms for
  spread-spectrum systems,'' \emph{IEEE Transactions on Vehicular Technology},
  vol.~59, no.~3, pp. 1217--1228, March 2010.

\bibitem{ccmjio2010}
L.~Wang, R.~C. de~Lamare, and M.~Yukawa, ``Adaptive reduced-rank constrained
  constant modulus algorithms based on joint iterative optimization of filters
  for beamforming,'' \emph{IEEE Transactions on Signal Processing}, vol.~58,
  no.~6, pp. 2983--2997, June 2010.

\bibitem{sjidf2010}
R.~Fa, R.~C. de~Lamare, and L.~Wang, ``Reduced-rank stap schemes for airborne
  radar based on switched joint interpolation, decimation and filtering
  algorithm,'' \emph{IEEE Transactions on Signal Processing}, vol.~58, no.~8,
  pp. 4182--4194, Aug 2010.

\bibitem{jiomimo2011}
R.~C. de~Lamare and R.~Sampaio-Neto, ``Adaptive reduced-rank equalization
  algorithms based on alternating optimization design techniques for mimo
  systems,'' \emph{IEEE Transactions on Vehicular Technology}, vol.~60, no.~6,
  pp. 2482--2494, July 2011.

\bibitem{jiostap2011}
R.~Fa and R.~C.~D. Lamare, ``Reduced-rank stap algorithms using joint iterative
  optimization of filters,'' \emph{IEEE Transactions on Aerospace and
  Electronic Systems}, vol.~47, no.~3, pp. 1668--1684, July 2011.

\bibitem{barc2011}
R.~C. de~Lamare, R.~Sampaio-Neto, and M.~Haardt, ``Blind adaptive constrained
  constant-modulus reduced-rank interference suppression algorithms based on
  interpolation and switched decimation,'' \emph{IEEE Transactions on Signal
  Processing}, vol.~59, no.~2, pp. 681--695, Feb 2011.

\bibitem{uwbccm2011}
S.~Li and R.~C. de~Lamare, ``Blind reduced-rank adaptive receivers for ds-uwb
  systems based on joint iterative optimization and the constrained constant
  modulus criterion,'' \emph{IEEE Transactions on Vehicular Technology},
  vol.~60, no.~6, pp. 2505--2518, July 2011.

\bibitem{wlmwf2012}
N.~Song, R.~C. de~Lamare, M.~Haardt, and M.~Wolf, ``Adaptive widely linear
  reduced-rank interference suppression based on the multistage wiener
  filter,'' \emph{IEEE Transactions on Signal Processing}, vol.~60, no.~8, pp.
  4003--4016, Aug 2012.

\bibitem{wljio2014}
N.~Song, W.~U. Alokozai, R.~C. de~Lamare, and M.~Haardt, ``Adaptive widely
  linear reduced-rank beamforming based on joint iterative optimization,''
  \emph{IEEE Signal Processing Letters}, vol.~21, no.~3, pp. 265--269, March
  2014.

\bibitem{dfjio2014}
L.~Wang, R.~C. de~Lamare, and M.~Haardt, ``Direction finding algorithms based
  on joint iterative subspace optimization,'' \emph{IEEE Transactions on
  Aerospace and Electronic Systems}, vol.~50, no.~4, pp. 2541--2553, October
  2014.

\bibitem{rdrb2015}
S.~D. Somasundaram, N.~H. Parsons, P.~Li, and R.~C. de~Lamare,
  ``Reduced-dimension robust capon beamforming using krylov-subspace
  techniques,'' \emph{IEEE Transactions on Aerospace and Electronic Systems},
  vol.~51, no.~1, pp. 270--289, January 2015.

\bibitem{mserjidf2015}
Y.~Cai, R.~C. de~Lamare, B.~Champagne, B.~Qin, and M.~Zhao, ``Adaptive
  reduced-rank receive processing based on minimum symbol-error-rate criterion
  for large-scale multiple-antenna systems,'' \emph{IEEE Transactions on
  Communications}, vol.~63, no.~11, pp. 4185--4201, Nov 2015.

\bibitem{dfalrd2016}
L.~Qiu, Y.~Cai, R.~C. de~Lamare, and M.~Zhao, ``Reduced-rank doa estimation
  algorithms based on alternating low-rank decomposition,'' \emph{IEEE Signal
  Processing Letters}, vol.~23, no.~5, pp. 565--569, May 2016.

\bibitem{GolubVanLoan96}
G.~H. Golub and C.~F. van Loan, \emph{{Matrix Computations}}, 3rd ed., Johns
  Hopkins Univ. Press, Baltimore, MD, (1996).

\bibitem{Chan87}
T.~F. Chan, ``{Rank revealing QR factorizations},'' \emph{Linear Algebra and
  its Applications}, vol. 88--89, pp. 67--82, Apr 1987.

\bibitem{GuEisenstat96}
M.~Gu and S.~C. Eisenstat, ``{Efficient Algorithms for Computing a Strong
  Rank-Revealing QR Factorization},'' \emph{SIAM J. Sci. Comput.}, vol.~17,
  no.~4, pp. 848–--869, 1996.

\bibitem{Stewart98}
G.~W. Stewart, \emph{{Matrix Algorithms: Volume 1: Basic Decompositions}},
  SIAM, Philadelphia, PA, (1998).

\bibitem{HMT2009}
N.~Halko, P.-G. Martinsson, and J.~A. Tropp, ``{Finding structure with
  randomness: Probabilistic algorithms for constructing approximate matrix
  decompositions},'' \emph{SIAM Review}, vol.~53, no.~2, pp. 217--288, Jun
  2011.

\bibitem{Gu2015}
M.~Gu, ``{Subspace Iteration Randomization and Singular Value Problems},''
  \emph{SIAM J. Sci. Comput.}, vol.~37, no.~3, pp. A1139--A1173.

\bibitem{FriezeKVS04}
A.~Frieze, R.~Kannan, and S.~Vempala, ``Fast monte-carlo algorithms for finding
  low-rank approximations,'' \emph{J. ACM}, vol.~51, no.~6, pp. 1025--1041,
  Nov. 2004.

\bibitem{Rokhlin09}
V.~Rokhlin, A.~Szlam, and M.~Tygert, ``{A randomized algorithm for principal
  component analysis},'' \emph{SIAM. J. Matrix Anal. \& Appl.}, vol.~31, no.~3,
  pp. 1100--1124, 2009.

\bibitem{DemmGHL12}
J.~Demmel, L.~Grigori, M.~Hoemmen, and J.~Langou, ``{Communication-optimal
  Parallel and Sequential QR and LU Factorizations},'' \emph{SIAM J. Sci.
  Comput.}, vol.~34, no.~1, pp. A206–--A239, 2012.

\bibitem{DemGGX15}
J.~Demmel, L.~Grigori, M.~Gu, and H.~Xiang, ``{Communication Avoiding Rank
  Revealing QR Factorization with Column Pivoting},'' \emph{SIAM J. Matrix
  Anal. \& Appl.}, vol.~36, no.~1, pp. 55--89, 2015.

\bibitem{DuerschGu2017}
J.~A. Duersch and M.~Gu, ``{Randomized QR with Column Pivoting},'' \emph{SIAM
  J. Sci. Comput.}, vol.~39, no.~4, pp. C263--C291, 2017.

\bibitem{CSPW2009}
V.~Chandrasekaran, S.~Sanghavi, P.~a. Parrilo, and A.~S. Willsky,
  ``{Rank-Sparsity Incoherence for Matrix Decomposition},'' \emph{SIAM J.
  Opt.}, vol.~21, no.~2, pp. 572--596, 2009.

\bibitem{CLMW2009}
E.~J. Cand{\`{e}}s, X.~Li, Y.~Ma, and J.~Wright, ``{Robust principal component
  analysis?}'' \emph{Journal of the ACM}, vol.~58, no.~3, pp. 1--37, May 2011.

\bibitem{CalvettiRS94}
D.~Calvetti, L.~Reichel, and D.~Sorensen, ``{An implicitly restarted Lanczos
  method for large symmetric eigenvalue problems},'' \emph{ETNA}, vol.~2, pp.
  1--21, 1994.

\bibitem{TrYUC17}
J.~Tropp, A.~Yurtsever, M.~Udell, and V.~Cevher, ``{Practical Sketching
  Algorithms for Low-Rank Matrix Approximation},'' \emph{SIAM J. Matrix Anal.
  \& Appl.}, vol.~38, no.~4, pp. 1454--1485, 2017.

\bibitem{MFKDeTSP18}
M.~F. Kaloorazi and R.~C. de~Lamare, ``{Subspace-Orbit Randomized Decomposition
  for Low-Rank Matrix Approximations},'' \emph{IEEE Trans. Signal Process.},
  vol.~66, no.~16, pp. 4409--4424, Aug 2018.

\bibitem{corutv}
M.~F. {Kaloorazi} and R.~C. {de Lamare}, ``Compressed randomized utv
  decompositions for low-rank matrix approximations,'' \emph{IEEE Journal of
  Selected Topics in Signal Processing}, vol.~12, no.~6, pp. 1155--1169, Dec
  2018.

\bibitem{Dongarra17}
J.~Dongarra, S.~Tomov, P.~Luszczek, J.~Kurzak, M.~Gates, I.~Yamazaki, H.~Anzt,
  A.~Haidar, and A.~Abdelfattah, ``{With Extreme Computing, the Rules Have
  Changed},'' \emph{Computing in Science and Engineering}, vol.~19, no.~3, pp.
  52--62, May 2017.

\bibitem{MartinssonHQRRP2017}
P.-G. Martinsson, G.~Quintana~Ort\'{i}, N.~Heavner, and R.~van~de Geijn,
  ``{Householder QR Factorization With Randomization for Column Pivoting
  (HQRRP)},'' \emph{SIAM J. Sci. Comput.}, vol.~39, no.~2, pp. C96--C115, 2017.

\bibitem{Bertsekas1982}
D.~Bertsekas, \emph{{Constrained Optimization and Lagrange Multiplier Method}},
  Academic Press, (1982).

\bibitem{LLS2011}
Z.~Lin, R.~Liu, and Z.~Su, ``{Linearized Alternating Direction Method with
  Adaptive Penalty for Low-Rank Representation},'' in \emph{NIPS}, no.~1, 2011,
  pp. 1--9.

\bibitem{Toh2010}
K.~C. Toh and S.~Yun, ``{An accelerated proximal gradient algorithm for nuclear
  norm regularized linear least squares problems},'' \emph{Pac. J. Optim.},
  vol.~6, no.~3, pp. 615--640, 2010.

\bibitem{Larsen98}
R.~M. Larsen, \emph{{Efficient algorithms for helioseismic inversion}}, PhD
  Thesis, University of Aarhus, Denmark (1998).

\bibitem{Hale2008}
E.~Hale, W.~Yin, and Y.~Zhang, ``{Fixed-Point Continuation for
  $\ell_1$-Minimization: Methodology and Convergence},'' \emph{SIAM J. Opt.},
  vol.~19, no.~3, pp. 1107–--1130, 2008.

\bibitem{FierroHanHan99}
R.~D. Fierro, P.~C. Hansen, and H.~P.~S. K., ``{UTV Tools: Matlab templates for
  rank-revealing UTV decompositions},'' \emph{Numerical Algorithms}, vol.~20,
  pp. 165--–194, 1999.

\bibitem{rstap2012}
Z.~Yang, R.~C. de~Lamare, and X.~Li, ``$l_1$-regularized stap algorithms with a
  generalized sidelobe canceler architecture for airborne radar,'' \emph{IEEE
  Transactions on Signal Processing}, vol.~60, no.~2, pp. 674--686, Feb 2012.

\bibitem{sastap2012}
------, ``Sparsity-aware space-time adaptive processing algorithms with
  $l_1$-norm regularisation for airborne radar,'' \emph{IET Signal Processing},
  vol.~6, no.~5, pp. 413--423, July 2012.

\bibitem{locsme2014}
H.~Ruan and R.~C. de~Lamare, ``Robust adaptive beamforming using a
  low-complexity shrinkage-based mismatch estimation algorithm,'' \emph{IEEE
  Signal Processing Letters}, vol.~21, no.~1, pp. 60--64, Jan 2014.

\bibitem{saalt2014}
R.~C. de~Lamare and R.~Sampaio-Neto, ``Sparsity-aware adaptive algorithms based
  on alternating optimization and shrinkage,'' \emph{IEEE Signal Processing
  Letters}, vol.~21, no.~2, pp. 225--229, Feb 2014.

\bibitem{als2015}
\BIBentryALTinterwordspacing
S.~Xu, R.~C. de~Lamare, and H.~V. Poor, ``Adaptive link selection algorithms
  for distributed estimation,'' \emph{EURASIP Journal on Advances in Signal
  Processing}, vol. 2015, no.~1, p.~86, Oct 2015. [Online]. Available:
  \url{https://doi.org/10.1186/s13634-015-0272-4}
\BIBentrySTDinterwordspacing

\bibitem{dce2015}
------, ``Distributed compressed estimation based on compressive sensing,''
  \emph{IEEE Signal Processing Letters}, vol.~22, no.~9, pp. 1311--1315, Sept
  2015.

\bibitem{damdc2016}
T.~G. Miller, S.~Xu, R.~C. de~Lamare, and H.~V. Poor, ``Distributed spectrum
  estimation based on alternating mixed discrete-continuous adaptation,''
  \emph{IEEE Signal Processing Letters}, vol.~23, no.~4, pp. 551--555, April
  2016.

\bibitem{okspme2016}
H.~Ruan and R.~C. de~Lamare, ``Robust adaptive beamforming based on low-rank
  and cross-correlation techniques,'' \emph{IEEE Transactions on Signal
  Processing}, vol.~64, no.~15, pp. 3919--3932, Aug 2016.

\bibitem{bfpeg2011}
A.~G.~D. Uchoa, C.~Healy, R.~C. de~Lamare, and R.~D. Souza, ``Design of ldpc
  codes based on progressive edge growth techniques for block fading
  channels,'' \emph{IEEE Communications Letters}, vol.~15, no.~11, pp.
  1221--1223, November 2011.

\bibitem{vfap2012}
J.~Liu and R.~C. de~Lamare, ``Low-latency reweighted belief propagation
  decoding for ldpc codes,'' \emph{IEEE Communications Letters}, vol.~16,
  no.~10, pp. 1660--1663, October 2012.

\bibitem{memd2016}
C.~T. Healy and R.~C. de~Lamare, ``Design of ldpc codes based on multipath emd
  strategies for progressive edge growth,'' \emph{IEEE Transactions on
  Communications}, vol.~64, no.~8, pp. 3208--3219, Aug 2016.

\bibitem{spa2008}
R.~C. de~Lamare and R.~Sampaio-Neto, ``Minimum mean-squared error iterative
  successive parallel arbitrated decision feedback detectors for ds-cdma
  systems,'' \emph{IEEE Transactions on Communications}, vol.~56, no.~5, pp.
  778--789, May 2008.

\bibitem{gbd2013}
K.~Zu, R.~C. de~Lamare, and M.~Haardt, ``Generalized design of low-complexity
  block diagonalization type precoding algorithms for multiuser mimo systems,''
  \emph{IEEE Transactions on Communications}, vol.~61, no.~10, pp. 4232--4242,
  October 2013.

\bibitem{tds2012}
P.~Clarke and R.~C. de~Lamare, ``Transmit diversity and relay selection
  algorithms for multirelay cooperative mimo systems,'' \emph{IEEE Transactions
  on Vehicular Technology}, vol.~61, no.~3, pp. 1084--1098, March 2012.

\bibitem{lsmimo2015}
W.~Zhang, H.~Ren, C.~Pan, M.~Chen, R.~C. de~Lamare, B.~Du, and J.~Dai,
  ``Large-scale antenna systems with ul/dl hardware mismatch: Achievable rates
  analysis and calibration,'' \emph{IEEE Transactions on Communications},
  vol.~63, no.~4, pp. 1216--1229, April 2015.

\bibitem{mfsic2011}
P.~Li, R.~C. de~Lamare, and R.~Fa, ``Multiple feedback successive interference
  cancellation detection for multiuser mimo systems,'' \emph{IEEE Transactions
  on Wireless Communications}, vol.~10, no.~8, pp. 2434--2439, August 2011.

\bibitem{mbthp2014}
K.~Zu, R.~C. de~Lamare, and M.~Haardt, ``Multi-branch tomlinson-harashima
  precoding design for mu-mimo systems: Theory and algorithms,'' \emph{IEEE
  Transactions on Communications}, vol.~62, no.~3, pp. 939--951, March 2014.

\bibitem{mbdf2013}
R.~C. de~Lamare, ``Adaptive and iterative multi-branch mmse decision feedback
  detection algorithms for multi-antenna systems,'' \emph{IEEE Transactions on
  Wireless Communications}, vol.~12, no.~10, pp. 5294--5308, October 2013.

\bibitem{mfdf2012}
P.~Li and R.~C.~D. Lamare, ``Adaptive decision-feedback detection with
  constellation constraints for mimo systems,'' \emph{IEEE Transactions on
  Vehicular Technology}, vol.~61, no.~2, pp. 853--859, Feb 2012.

\bibitem{armo2013}
T.~Peng, R.~C. de~Lamare, and A.~Schmeink, ``Adaptive distributed space-time
  coding based on adjustable code matrices for cooperative mimo relaying
  systems,'' \emph{IEEE Transactions on Communications}, vol.~61, no.~7, pp.
  2692--2703, July 2013.

\bibitem{sicdma2011}
Y.~Cai, R.~C. d.~Lamare, and R.~Fa, ``Switched interleaving techniques with
  limited feedback for interference mitigation in ds-cdma systems,'' \emph{IEEE
  Transactions on Communications}, vol.~59, no.~7, pp. 1946--1956, July 2011.

\bibitem{mbsic2011}
R.~Fa and R.~C.~D. Lamare, ``Multi-branch successive interference cancellation
  for mimo spatial multiplexing systems: Design, analysis and adaptive
  implementation,'' \emph{IET Communications}, vol.~5, no.~4, pp. 484--494,
  March 2011.

\bibitem{bfidd2016}
A.~G.~D. Uchoa, C.~T. Healy, and R.~C. de~Lamare, ``Iterative detection and
  decoding algorithms for mimo systems in block-fading channels using ldpc
  codes,'' \emph{IEEE Transactions on Vehicular Technology}, vol.~65, no.~4,
  pp. 2735--2741, April 2016.

\bibitem{rmbthp2014}
L.~Zhang, Y.~Cai, R.~C. de~Lamare, and M.~Zhao, ``Robust multibranch
  tomlinson-harashima precoding design in amplify-and-forward mimo relay
  systems,'' \emph{IEEE Transactions on Communications}, vol.~62, no.~10, pp.
  3476--3490, Oct 2014.

\bibitem{did2014}
P.~Li and R.~C. de~Lamare, ``Distributed iterative detection with reduced
  message passing for networked mimo cellular systems,'' \emph{IEEE
  Transactions on Vehicular Technology}, vol.~63, no.~6, pp. 2947--2954, July
  2014.

\bibitem{jpais2012}
R.~C.~D. Lamare, ``Joint iterative power allocation and linear interference
  suppression algorithms for cooperative ds-cdma networks,'' \emph{IET
  Communications}, vol.~6, no.~13, pp. 1930--1942, Sept 2012.

\bibitem{badstbc2016}
T.~Peng and R.~C. de~Lamare, ``Adaptive buffer-aided distributed space-time
  coding for cooperative wireless networks,'' \emph{IEEE Transactions on
  Communications}, vol.~64, no.~5, pp. 1888--1900, May 2016.

\bibitem{wlbd2017}
W.~Zhang, R.~C. de~Lamare, C.~Pan, M.~Chen, J.~Dai, B.~Wu, and X.~Bao, ``Widely
  linear precoding for large-scale mimo with iqi: Algorithms and performance
  analysis,'' \emph{IEEE Transactions on Wireless Communications}, vol.~16,
  no.~5, pp. 3298--3312, May 2017.

\bibitem{mmimo2013}
R.~C. de~Lamare, ``Massive mimo systems: Signal processing challenges and
  future trends,'' \emph{URSI Radio Science Bulletin}, vol. 2013, no. 347, pp.
  8--20, Dec 2013.

\bibitem{bbprec2017}
L.~T.~N. Landau and R.~C. de~Lamare, ``Branch-and-bound precoding for multiuser
  mimo systems with 1-bit quantization,'' \emph{IEEE Wireless Communications
  Letters}, vol.~6, no.~6, pp. 770--773, Dec 2017.

\bibitem{1bitidd2018}
Z.~Shao, R.~C. de~Lamare, and L.~T.~N. Landau, ``Iterative detection and
  decoding for large-scale multiple-antenna systems with 1-bit adcs,''
  \emph{IEEE Wireless Communications Letters}, vol.~7, no.~3, pp. 476--479,
  June 2018.

\bibitem{baplnc2018}
J.~Gu, R.~C. de~Lamare, and M.~Huemer, ``Buffer-aided physical-layer network
  coding with optimal linear code designs for cooperative networks,''
  \emph{IEEE Transactions on Communications}, vol.~66, no.~6, pp. 2560--2575,
  June 2018.

\end{thebibliography}
\end{document}